\documentclass[12pt]{article}
\usepackage[dvips]{graphicx}
\DeclareGraphicsExtensions{.pdf,.jpg}

\usepackage{amsfonts,amsmath,amsthm,amssymb}
\usepackage{mathrsfs}
\usepackage[onehalfspacing]{setspace}

\author{Leonardo Ort\'{i}z\footnote{leonardo.ortiz@correo.nucleares.unam.mx}\\
Instituto de Ciencias Nucleares, UNAM\\
M\'{e}xico, D. F. 04510, M\'{e}xico}

\title{Notes on quantum fields on two dimensional spacetimes}

\begin{document}

\maketitle

\begin{center}
\small{\textbf{Abstract}}
\end{center}
\small{We point out how to construct the Hartle-Hawking-Israel state for the minimaly coupled massless quantum real scalar field in the two dimensional BTZ black hole. We also calculate the renormalized energy-momentum tensor for the same field in the eternal CGHS black hole, AdS, Robertson-Walker and Rindler spacetime in two dimensions. We also discuss the Boulware, the Hartle-Hawking-Israel and the Unruh state for the eternal CGHS black hole.}\\

\normalsize{Keywords: Quantum Fields, BTZ black hole, CGHS black hole, AdS spacetime, Robertson-Walker spacetime, Rindler spacetime, energy-momentum tensor.}\\

PACS: 04.62.+v, 04.70.Dy
\newpage

\section{Introduction} 

The study of quantum fields on a fixed background has been very important in understanding some aspects which the still elusive theory of quantum gravity would give a complete account. Since the discovery of the Hawking effect, circa 1975, a lot of work has been done on different backgrounds, in particular on black holes. Perhaps for their relevance for our world the study of quantum fields on four dimensional spacetimes has gotten a lot of atention. However from the theoretical point of view the study of quantum fields on two dimensional backgrounds has an intrinsic interest too. The purpose of these notes is to study some aspects of quantum fields in several two dimensional backgrounds. First we point out how to construct the Hartle-Hawking-Israel state for the real massless scalar feld in the two dimensional BTZ black hole. Then we study the renormalized energy-momentum tensor for the same field in the following two dimensional backgrounds: CGHS, AdS, Robertson-Walker, and Rindler. In passing we discuss the Boulware, the Hartle-Hawking-Israel and the Unruh state for the eternal CGHS black hole. Although a lot of work has been done on quantum fields on two dimensional spacetimes as far as we know the aspects which we study in this work have not been considered before. Hence our work fills a gap in the literature.

The organization of the paper is the following: in section 2 we point out how to construct the Hartle-Hawking-Israel state, in section 3 we calculate the renormalized energy-momentum tensor in the CGHS model \cite{CGHS92}, the renormalized energy-momentum tensor in two dimensional AdS, Robertson-Walker and Rindler spacetime, finally in section 4 we give some final comments on this work.

\section{The Hartle-Hawking-Israel state}

The two dimensional BTZ black hole can be considered as a solution of the JT theory \cite{mCasMi95}, its metric is given by\footnote{This metric can also be considered as a dimensional reduction of the BTZ metric in three dimensions \cite{aAchumOrt93}.}
\begin{equation}
ds^2=-\left(-M+\frac{r^2}{l^2}\right)dt^2+\frac{dr^2}{\left(-M+\frac{r^2}{l^2}\right)},
\end{equation}
where $M$ and $l$ are parameters of the theory. This metric has a Kruskal like extension \cite{lOr11} and we can introduce a tortoise like coordinate \cite{lOr11} $r^{*}$ in such a way that the metric takes the form
\begin{equation}\label{eq.1}
ds^2=\left(-M+\frac{r^2}{l^2}\right)(dt^2+dx^2),
\end{equation}
where we have made $r^{*}=x$ and $r$ is an implicit function of $x$. The coordinate $x$ is $-\infty$ at the horizon and $0$ at infinity.

Now we consider the massless conformally coupled real scalar field with equation of motion
\begin{equation}\label{eq.2}
\nabla_{\mu}\nabla^{\mu}\phi=0
\end{equation}
in the metric (\ref{eq.1}). The equation (\ref{eq.2}) is such that it reduces to
\begin{equation}
(-\partial^2_{t}+\partial^2_{x})\phi=0.
\end{equation}
If we assume a harmonic dependence in time and impose Dirichlet boundary conditions at infinity\footnote{We have to impose boundary conditions at infinity since the BTZ black hole has a time-like boundary at infinity.} then we have the solutions
\begin{equation}\label{eq.3}
\phi(t,x)=\frac{1}{\sqrt{\pi\omega}}e^{-i\omega t}\sin(\omega x).
\end{equation}
These solutions are normalized with respect to the Klein-Gordon inner product \footnote{If we choose Neumann boundary conditions then the normalized solutions are with cosine instead of sine.}.

The Kruskal coordinates $T$ and $R$ are given in
    terms of $t$ and $r^{\ast}$ as \cite{lOr11}:
    
    In $\mathcal{R}_{K}$
    \begin{equation}\label{E:b28}
    T=e^{\frac{r_{+}}{l^{2}}r^{\ast}}\sinh\left(\frac{r_{+}}{l^{2}}t\right)\qquad
    R=e^{\frac{r_{+}}{l^{2}}r^{\ast}}\cosh\left(\frac{r_{+}}{l^{2}}t\right).
    \end{equation}
    In $\mathcal{L}_{K}$
    \begin{equation}\label{E:b28a}
    T=-e^{\frac{r_{+}}{l^{2}}r^{\ast}}\sinh\left(\frac{r_{+}}{l^{2}}t\right)\qquad
    R=-e^{\frac{r_{+}}{l^{2}}r^{\ast}}\cosh\left(\frac{r_{+}}{l^{2}}t\right).
    \end{equation}
    In $\mathcal{F}_{K}$
    \begin{equation}\label{E:b28b}
    T=e^{\frac{r_{+}}{l^{2}}r^{\ast}}\cosh\left(\frac{r_{+}}{l^{2}}t\right)\qquad
    R=e^{\frac{r_{+}}{l^{2}}r^{\ast}}\sinh\left(\frac{r_{+}}{l^{2}}t\right).
    \end{equation}
    In $\mathcal{P}_{K}$
    \begin{equation}\label{E:b28c}
    T=-e^{\frac{r_{+}}{l^{2}}r^{\ast}}\cosh\left(\frac{r_{+}}{l^{2}}t\right)\qquad
    R=-e^{\frac{r_{+}}{l^{2}}r^{\ast}}\sinh\left(\frac{r_{+}}{l^{2}}t\right).
    \end{equation}
    Here $\mathcal{R}_{K}$, $\mathcal{L}_{K}$, $\mathcal{F}_{K}$ and $\mathcal{P}_{K}$ are the right, left, future and past regions of the Kruskal spacetime. From these expressions we see that the $(T,R)$ and $(t,r^{\ast})$ coordinates
    are related as the Minkowski and Rindler coordinates in flat
    spacetime are.
    
    Following \cite{cKr10} we can introduce Rindler like modes in $\mathcal{R}_{K}$ and $\mathcal{L}_{K}$. These modes will be given in terms of (\ref{eq.3}). It is clear from the expressions of the normalized modes and the form of the Kruskal coordiantes in terms of $t$ and $r^{\ast}$ that every step given in \cite{cKr10} applies to the present case and then we can obtain the Hartle-Hawking-Israel state which would be the analogous of the Minkowski vacuum in terms of Rindler vacuum. The state we obtain is analogous to the one obtained by \cite{wIsr76} however in our case we have given an explicit expression for the functions which define the Rindler like quantization whereas Israel just assumed the existence of such functions. Also we point out that our arguments clarify the way the Hartle-Hawking-Israel state is obtained from first principles.

\section{The energy-momentum tensor}

It is well-known that the energy-momentum tensor in Quantum Field
Theory in curved spacetime is a subtle issue. This is principally
due to the divergences which occur when the expectation value of
it in a certain state is calculated, see for example
\cite{ndBpcwD82} for an extensive discussion. However, since it
contains important physical information of the field it is worth
trying to calculate it. It turns out that in 1+1 dimensions most
of the difficulties can be removed and it is possible to obtain
closed expressions for it \cite{pDav77}. In this paper we will
exploit this fact and will calculate this quantity for several two dimensional spacetimes. Even though a great amount of work on
calculating the expectation value of the energy-momentum tensor in
several two dimensional spacetimes has been done, see for example
references in \cite{pDav77}, \cite{rBalaFab98} and \cite{aFabjNs05}, as far as we
know the study of this tensor in the CGHS model using the methods described in \cite{pDav77} has not been done before. One of the main motivation of this work is to fill this
gap in the literature. Also it is worth mentioning that although
the calculations are very simple interesting results are obtained
and closed expressions as well.

\subsection{The energy-momentum tensor in 1+1 dimensions}

In 1+1 dimensions the energy-momentum tensor is almost determined
by its trace. In what follows we give the basic formule for
calculating this quantity.

Let us consider the following metric
\begin{equation}\label{E:1e6}
ds^{2}=C(-dt^{2}+dx^{2})=-Cdudv,
\end{equation}
where $u=t-x$ and $v=t+x$. Since every 1+1 dimensional metric is
conformal to a 1+1 dimensional metric in Minkowski spacetime, the
metric (\ref{E:1e6}) is very general. The function $C$ in general
depends on both variables in the metric. In these circumstances
the expectation value of the trace of the energy-momentum is
\cite{pDav77}
\begin{equation}\label{E:2e6}
<T^{\mu}_{\mu}>=-\frac{R}{24\pi}=\frac{1}{6\pi}\left(\frac{C_{uv}}{C^{2}}-\frac{C_{u}C_{v}}{C^{3}}\right),
\end{equation}
where $R$ is the Ricci scalar and $C_{u}=\frac{\partial}{\partial
u}C$, etc. The last expression holds for the real scalar field.
The expectation value of the components of the energy-momentum
tensor in null coordinates is
\begin{equation}\label{E:3e6}
\langle
T_{uu}\rangle=-\frac{1}{12\pi}C^{1/2}\partial_{u}^{2}C^{-1/2}+f(u)
\end{equation}
\begin{equation}\label{E:4e6}
\langle
T_{vv}\rangle=-\frac{1}{12\pi}C^{1/2}\partial_{v}^{2}C^{-1/2}+g(v)
\end{equation}
where $f$ and $g$ are arbitrary functions of $u$ and $v$
respectively. These functions contain information about the state
with respect to which the expectation value is taken. The mixed
components are given by
\begin{equation}\label{E:5e6}
\langle T_{uv}\rangle=-\frac{CR}{96\pi}.
\end{equation}
Now let us apply these formulae to the eternal CGHS black hole.

\subsection{The energy-momentum tensor in the eternal CGHS black hole}

The metric for the CGHS black hole can
be written in the form (\ref{E:1e6}) with \cite{sbGiwmNe92}
\begin{equation}\label{E:6e6}
C=\left(1+\frac{M}{\lambda}e^{\lambda(u-v)}\right)^{1/2},
\end{equation}
where $M$ is the mass of the black hole and $\lambda$ is a constant of the theory.

Using the previous expression for $C$ we obtain
\begin{equation}\label{E:7e6}
\langle T_{uu}\rangle=-\frac{M}{24\pi}\left[\frac{\lambda e^{\lambda(u-v)}}{1+\frac{M}{\lambda}e^{\lambda(u-v)}}-\frac{M}{2}\frac{e^{2\lambda(u-v)}}{(1+\frac{M}{\lambda}e^{\lambda(u-v)})^2}\right]+f(u)
\end{equation}
and
\begin{equation}\label{E:8e6}
\langle T_{vv}\rangle=-\frac{M}{24\pi}\left[\frac{\lambda e^{\lambda(u-v)}}{1+\frac{M}{\lambda}e^{\lambda(u-v)}}-\frac{M}{2}\frac{e^{2\lambda(u-v)}}{(1+\frac{M}{\lambda}e^{\lambda(u-v)})^2}\right]+g(v).
\end{equation}

If we choose $f(u)=g(v)=0$ we obtain the analogue of the Boulware
state in Schwarzschild spacetime which is singular at the horizon. This can be seen as follows. With this choice, at the horizon, we have
\begin{equation}
\langle T_{uu}\rangle=\langle T_{vv}\rangle=-\frac{1}{48\pi}\lambda^{2}.
\end{equation}
Then in Kruskal null coordinates $-\lambda U=e^{-\lambda u}$ and $\lambda V=e^{\lambda v}$ we have
\begin{equation}
\langle T_{UU}\rangle=\frac{1}{U^{2}}\langle T_{uu}\rangle,\hspace{0.5cm} \langle T_{VV}\rangle=\frac{1}{V^{2}}\langle T_{vv}\rangle
\end{equation}
which clearly diverge at the horizon $U=V=0$.
We can also obtain the analogous of the
Hartle-Hawking state which is regular in both the future and the
past horizons. The value of $f(u)$ and $g(v)$ can be obtained if
we express the energy-momentum tensor in Kruskal like coordinates,
U and V. In these coordinates the energy momentum tensor is
\begin{equation}\label{E:19e6}
\langle T_{UU}\rangle=\frac{1}{U^{2}}\langle
T_{uu}\rangle,
\end{equation}
\begin{equation}\label{E:20e6}
\langle T_{VV}\rangle=\frac{1}{V^{2}}\langle
T_{vv}\rangle,
\end{equation}
and
\begin{equation}\label{E:21e6}
\langle T_{UV}\rangle=0.
\end{equation}
Hence we demand that $f(u)=g(v)=\frac{1}{48\pi}\lambda^{2}$. And naturally the Unruh state is obtained by choosing $f(u)=\frac{1}{48\pi}\lambda^{2}$ and $g(v)=0$. With this choice we have
\begin{equation}
\langle T_{uu}\rangle=\frac{\lambda^{2}}{48\pi}\frac{1}{(1+\frac{M}{\lambda}e^{-\lambda(v-u)})^{2}}
\end{equation}
and
\begin{equation}
\langle T_{vv}\rangle=\frac{\lambda^{2}}{48\pi}\left[\frac{1}{(1+\frac{M}{\lambda}e^{-\lambda(v-u)})^{2}}-1\right].
\end{equation}
This state shows that $\langle T_{uu}\rangle$ is regular at future horizon and future infinity. Moreover it also shows a Hawking flux at future infinity. So it mimic the quantum state of a field living on a collapsing black hole. It is also easy to see that $\langle T_{vv}\rangle$ is singular at the past horizon.

\subsection{The energy-momentum tensor in AdS}

In AdS spacetime the metric is
\begin{equation}
ds^2=l^2\sec^2\rho(-dt^2+d\rho^2)=-l^2\sec^2\rho dudv,
\end{equation}
where $u=t-\rho$ and $v=t+\rho$. Then in this case $C=l^2\sec^2\left(\frac{v-u}{2}\right)$. And then we have
\begin{equation}
\partial_{u}^2C^{-\frac{1}{2}}=-\frac{1}{4}C^{-\frac{1}{2}},
\end{equation}
\begin{equation}
\partial_{v}^2C^{-\frac{1}{2}}=-\frac{1}{4}C^{-\frac{1}{2}}.
\end{equation}
From these equations then the renormalized energy-momentum tensor is
\begin{equation}
\left\langle T_{uu}\right\rangle=\frac{1}{48\pi}+f(u), 
\end{equation}
\begin{equation}
\left\langle T_{vv}\right\rangle=\frac{1}{48\pi}+g(v). 
\end{equation}
Here we should choose $f(u)$ and $g(v)$ equal to $-\frac{1}{48\pi}$ to be in agreement with the equation \cite{ndBpcwD82}
\begin{equation}
\left\langle T_{\mu\nu}\right\rangle=\frac{1}{2}g_{\mu\nu}T, 
\end{equation}
which is valid for AdS and $T$ is the trace of the renormalized energy-momentum tensor.

\subsection{The energy-momentum tensor in Robertson-Walker}

In this spacetime we will take our metric as follows \cite{ndBpcwD82}
\begin{equation}
ds^2=B\tanh \rho\eta(-d^2\eta+d^2x)=-B\tanh \rho\eta(dudv).
\end{equation}
This metric represents an universe expanding for some time. Then in the very early time we expect no particles, whereas in the late time limit we expect some particles created during the expanding time. Here $B$ and $\rho$ are constants. In this case our $C=B\tanh \rho\eta$.

Doing the calculation we find
\begin{equation}
C^{\frac{1}{2}}\partial_{u}^{2}C^{-\frac{1}{2}}=-\frac{\rho^2}{8\sinh^2\left(\frac{u+v}{2}\right)}\left(-4+\frac{1}{2}\frac{1}{\cosh^2\left(\frac{u+v}{2}\right)}\right)=C^{\frac{1}{2}}\partial_{v}^{2}C^{-\frac{1}{2}}.
\end{equation}
From this expresion we see that at very early and very late times we have $\left\langle T_{uu}\right\rangle =\left\langle T_{vv}\right\rangle =0$ if we choose $f(u)=g(v)=0$. Hence in this case the production of particles is encoded in the cross term
\begin{equation}
\langle T_{uv}\rangle=-\frac{CR}{96\pi}.
\end{equation}
Hence we choose our vacuum as having zero particles at very early time and with particles at late time described by the function $C$ and the Ricci scalar.

\subsection{The energy-momentum tensor in Rindler}

In this case our metric will be
\begin{equation}
ds^2=e^{2ax}(-d\eta^2+dx^2),
\end{equation}
where $a$ is the acceleration. Then in this case $C=e^{a(v-u)}$, with $u=\eta-x$ and $v=\eta+x$. After a little calculation we get
\begin{equation}
C^{\frac{1}{2}}\partial_{u}^{2}C^{-\frac{1}{2}}=\frac{a^2}{4},
\end{equation}
\begin{equation}
C^{\frac{1}{2}}\partial_{v}^{2}C^{-\frac{1}{2}}=\frac{a^2}{4}.
\end{equation}
Hence if we want to have the energy-momentum tensor in the Rindler vacuum we must choose $f(u)=g(v)=\frac{a^2}{48\pi}$ and then all the components of the energy-momentum tensor are zero.

\section{Final comments}

In this work we have discussed how to construct the Hartle-Hawking-Israel vacuum for the BTZ black hole in two dimensions. We have given the explicit modes to contruct this vacuum. We also have discussed the renormalized energy-momentum tensor in several spacetimes and found simple and closed expressions for it. We belive that this work fills a gap in the literature regarding quatum fields in two dimensions.\\\vspace{0.5cm}

\textbf{Acknowledgements}: We thank Sujoy K. Modak for stimulating discussions. This work was supported with DGAPA-UNAM postdoctoral fellowship.

\end{document}